\documentclass[12pt]{iopart} 
\usepackage{graphicx,color} 

\newcommand{\pq}[1]{\left[{#1}\right]}

\begin{document}

\title[]{Influence of rotational force fields on the determination of
the work done on a driven Brownian particle}

\author{Giuseppe Pesce$^{1,2}$, Giovanni Volpe$^{3,4}$, Alberto Imparato$^5$, 
Giulia Rusciano$^{1,2}$ and Antonio Sasso$^{1,2}$} 
\address{$^1$ Dipartimento di Scienze Fisiche, Universit`a di Napoli ''Federico II'', 
Complesso Universitario Monte S. Angelo Via Cintia, 80126 Napoli, Italy, EU} 
\address{$^2$ CNISM, Consorzio Nazionale Interuniversitario per le Scienze 
Fisiche della Materia, Sede di Napoli Napoli, Italy, EU}
\address{$^3$ Max-Planck-Institut f\"ur Metallforschung, Heisenbergstr. 3, 70569 
Stuttgart, Germany, EU} 
\address{$^4$ 2. Physikalisches Institut, Universit\"at Stuttgart, 
Pfaffenwaldring 57, 70569 Stuttgart, Germany, EU}
\address{$^5$ Department of Physics and Astronomy, University of Aarhus, 
Ny Munkegade, Building 1520, DK-8000 Aarhus C, Denmark, EU}
\ead{giuseppe.pesce@na.infn.it}

\begin{abstract} For a Brownian system the evolution of thermodynamic
quantities is a stochastic process.  In particular, the work performed
on a driven colloidal particle held in an optical trap changes for
each realization of the experimental manipulation, even though the
manipulation protocol remains unchanged.  Nevertheless, the work
distribution is governed by established laws.  Here, we show how
the measurement of the work distribution is influenced by the presence
of rotational, i.e. nonconservative, radiation forces.  Experiments
on particles of different materials show that the rotational radiation
forces, and therefore their effect on the work distributions, increase
with the particle refractive index.
\end{abstract}

\vspace{2pc}
\submitto{\JO}
\maketitle

\section{Introduction}

Most physical, chemical and biological phenomena present intrinsic
degrees of randomness. Such randomness can be treated within the
framework of statistical physics and thermodynamics. Equilibrium
thermodynamics, in particular, is a mature discipline with
well-established laws \cite{Gibbs,thermodynamics}. However, most
natural and engineered processes occur far from equilibrium, e.g., 
biomolecular reactions in molecular machines \cite{Wang1998}, 
oscillations of pumped mesoscopic chemical reaction systems \cite{Qian2002} 
and fast switching between phonon distributions in optical cavities
\cite{Chow2002}. Such processes cannot be treated within the framework
of classical equilibrium thermodynamics. This fact has motivated an
intense activity during the last years aiming at extending the results of
thermodynamics to out-of-equilibrium systems
\cite{mazur,Evans1993,Jarzynski1997,Crooks1998,Hummer2001,Reguera2005,
Imparato05,Imparato05b,Imparato06,ImparatoPeliti07,Kjelstrup}.

In this context, the study of thermodynamic quantities in microscopic
systems, such as biomolecules and nanomachines, has posed new
challenges due to the ineluctable presence of a Brownian noise
background, which prevents one from straightforwardly scaling down the
approaches employed on macroscopic systems.  For example, the presence of 
Brownian noise alters the measurement of forces acting on microscopic
objects, leading to artifacts if not correctly taken into
account \cite{Volpe2010PRL}.  Also the evolution of thermodynamics
quantities, e.g., work and entropy, although deterministic in
macroscopic systems, becomes stochastic in the microscopic
realm. Indeed, a different value is measured in each experimental
realization; even phenomena that are forbidden in a macroscopic
systems become possible at a microscopic level, e.g., entropy-decreasing
trajectories \cite{Wang2002,Imparato2007}.
Nevertheless, the probability distributions of these thermodynamics
quantities obey some deterministic laws
\cite{Imparato2007,maz_jarz_99,spe_seif_05,zon_cohen_04,tani_cohen_07}.

Optically trapped particles \cite{Ashkin1997} have emerged as a 
powerful model system to address experimentally novel concepts in the
context of statistical physics in a convenient way relying both on
the presence of a natural noisy background and on a finely
controllable deterministic optical force field \cite{Babic2005}.  Indeed, a
colloidal particle, i.e. a microscopic particle suspended in a fluid, is
set in constant movement by the presence of thermal fluctuations,
which are responsible for its diffusion and
introduce a well-defined noisy background. Furthermore, it is possible
to make use of optical forces to introduce deterministic perturbations
acting on the particle in a controllable way.  For example,
optically trapped particles have been used to study
experimentally many non-trivial, sometimes constructive, aspects of the
presence of noise, e.g., stochastic resonance
\cite{Gammaitoni1998}, stochastic activation \cite{Doering1992},
Brownian ratchets \cite{Astumian1997}, or stochastic resonant damping
\cite{Volpe2008}.

An implicit assumption in these studies is that the optical trap acts
as a harmonic potential, i.e. the restoring optical force is
proportional to the particle displacement from the trap center.
This entails that the optical force field should be conservative, excluding the possibility 
of a rotational component. This is actually true to a great extent in the $xy$-plane 
perpendicular to the beam propagation direction, at
least for a standard optical trap generated by a Gaussian
beam. However, due to the presence of the scattering force, this has
been shown not to be true in a plane parallel to the beam propagation,
e.g., $xz$-plane \cite{Ashkin1992,Merenda2006,Roichman2008,Pesce2009EPL}. For most
experimental situations the nonconservative effects are effectively
negligible, becoming significant only when the particle is allowed to
explore a large region of the optical field \cite{Pesce2009EPL}.  This
typically occurs for extremely low-power trapping, far away from the
trapping regimes usually used in experiments, or for particles driven
out-of-equilibrium.

In this article, we show paradigmatically how the determination of a
thermodynamic quantity, namely the work, is affected by the presence
of a rotational, i.e. nonconservative, force field.  We measure the
work performed by a fluid flow on three optically trapped colloidal
particles of comparable dimension, but different refractive index. 
For a larger refractive index, the rotational optical force field component increases
and, therefore, introduces larger artifacts into the measured work distributions.

\section{Experimental setup}

The experimental setup consists of an optical tweezers
build on a home-made optical microscope with a high-numerical-aperture
water-immersion objective lens (Olympus, UPLAPO60XW3, NA=1.2); the
optical trap is generated by a frequency and amplitude stabilized
Nd-YAG laser ($\mathrm{\lambda=1.064\,\mu m}$, $\mathrm{500\,mW}$
maximum output power, Innolight Mephisto)\cite{PesceRSI05}.

We use colloidal spheres of comparable diameter, i.e. $\sim 1 \, \mathrm{\mu m}$,
of three different material -- silica (Si), polystyrene (Ps) and
melamine (Me). The main parameters of the experimental setup and of the beads used in 
this experiment are reported in table \ref{table}.  The particles are diluted in
distilled deionized water to a final concentration of a few
particles/$\mathrm{\mu l}$. The sample cell is made with a $150 \, \mathrm{\mu
m}$-thick coverslip and a microscope slide, which are separated by a
$100\, \mathrm{\mu m}$-thick Parafilm spacer and sealed with vacuum grease to
prevent evaporation and contamination. Such sample cell is mounted on
a closed-loop piezoelectric stage (Physik Instrumente PI-517.3CL),
which allows movements with nanometer resolution. Obviously, a movement of the stage in
a given direction corresponds to a movement the optical trap focus in the
opposite direction from the perspective of the trapped particle. The sample temperature is continuously monitored
using a calibrated NTC thermistor positioned on the top surface of the
microscope slide and remained constant within $0.5\, \mathrm{K}$ during each
set of measurements.

A colloidal sphere is trapped and positioned in the middle of the sample
cell, i.e. far away from the surfaces to avoid spurious effects
arising from diffusion gradients \cite{Volpe2010PRL,Berg-Sorensen2004}.
Its $x$, $y$ and $z$ coordinates (figure \ref{fig1}(a)) are monitored
through the forward scattered light imaged at the back focal plane of
the condenser lens on a InGaAs Quadrant Photodiode (QPD, Hamamatsu
G6849) \cite{GittesOL98,BuoscioloOC04}. A digital oscilloscope
(Tektronix TDS5034B) is used for data-acquisition.  The QPD-response
is linear for displacements up to $\mathrm{300\, nm}$ ($\mathrm{2\,
  nm}$ resolution, $\mathrm{250\, kHz}$ bandwidth). The conversion
factor from voltage to distance is calibrated using the power spectral
density (PSD) method \cite{Berg-Sorensen2004}. 

To minimize external noise, the experimental setup is
mounted on a passive vibration isolation optical table.  Moreover, the
laser paths are kept as short as possible to avoid pointing
fluctuations and the laser beam is enclosed in plastic pipes wherever
possible.  Finally, all the setup is enclosed by a polystyrene box to
prevent air circulation and temperature drifts.  In order to check the
achieved stability, in figure \ref{fig2} the PSD for the $x$ (blue
squares) and $z$ (red diamonds) coordinates are presented for the Ps
particle. They fit well the expected theoretical curves (lines) in the range between 
$50 \, \mathrm{mHz}$ and $500\, \mathrm{Hz}$. In particular the PSDs show a 
very flat plateau up to the minimum frequency measured which corresponds to a period of $20\,\mathrm{s}$. This 
allows us to exclude the presence of low frequency noise, since, as described above,
each single measurement is performed within $10\,\mathrm{s}$.

\begin{table}
\caption{Particle and optical trap parameters. $k_{\rho}$ is the radial trap stiffness, $k_z$ is the radial trap stiffness, $\eta$ is the ratio $k_{z}/k_{\rho}$, $\eta$ their ratio and $\epsilon$ the relative contribution of the nonconservative force fields \cite{Pesce2009EPL}.} 
\label{table}
\begin{center}
\begin{tabular}{cccc} 
\hline \\
~                            & Silica         & Polystyrene   & Melamine \\ \hline
diameter ($\mathrm{\mu m}$)  & 0.97$\pm$0.01  & 0.99$\pm$0.02 & 1.00$\pm$0.02 \\
refractive index             & 1.37           & 1.59          & 1.68 \\ 
density ($\mathrm{g/cm^3}$)  & 1.96           & 1.06          & 1.51 \\ 
$k_{\rho}$ ~($\mathrm{pN/\mu m}$) & 1.38$\pm$0.09  & 1.17$\pm$0.07 & 0.84$\pm$0.08 \\ 
$k_z$ ~($\mathrm{pN/\mu m}$) & 0.26$\pm$0.01  & 0.27$\pm$0.01 & 0.27$\pm$0.01\\ 
$\eta$     & 0.19$\pm$0.01      & 0.23$\pm$0.01     & 0.32$\pm$0.01 \\
$\epsilon$ (\%)              & --             & 2.2$\pm$0.1   & 6.2$\pm$0.2 \\
laser power ($\mathrm{mW}$)  & $0.5\pm0.02$   & $0.4\pm0.02$  & $0.4\pm0.02$ \\
\hline
\end{tabular}
\end{center}
\end{table}

%$k_z$~($\mathrm{pN/\mu m}$)  & 0.26$\pm$0.09  & 0.27$\pm$     & 0.27 \\ 

\section{Manipulation protocol}

\begin{figure}
\includegraphics[width=14cm]{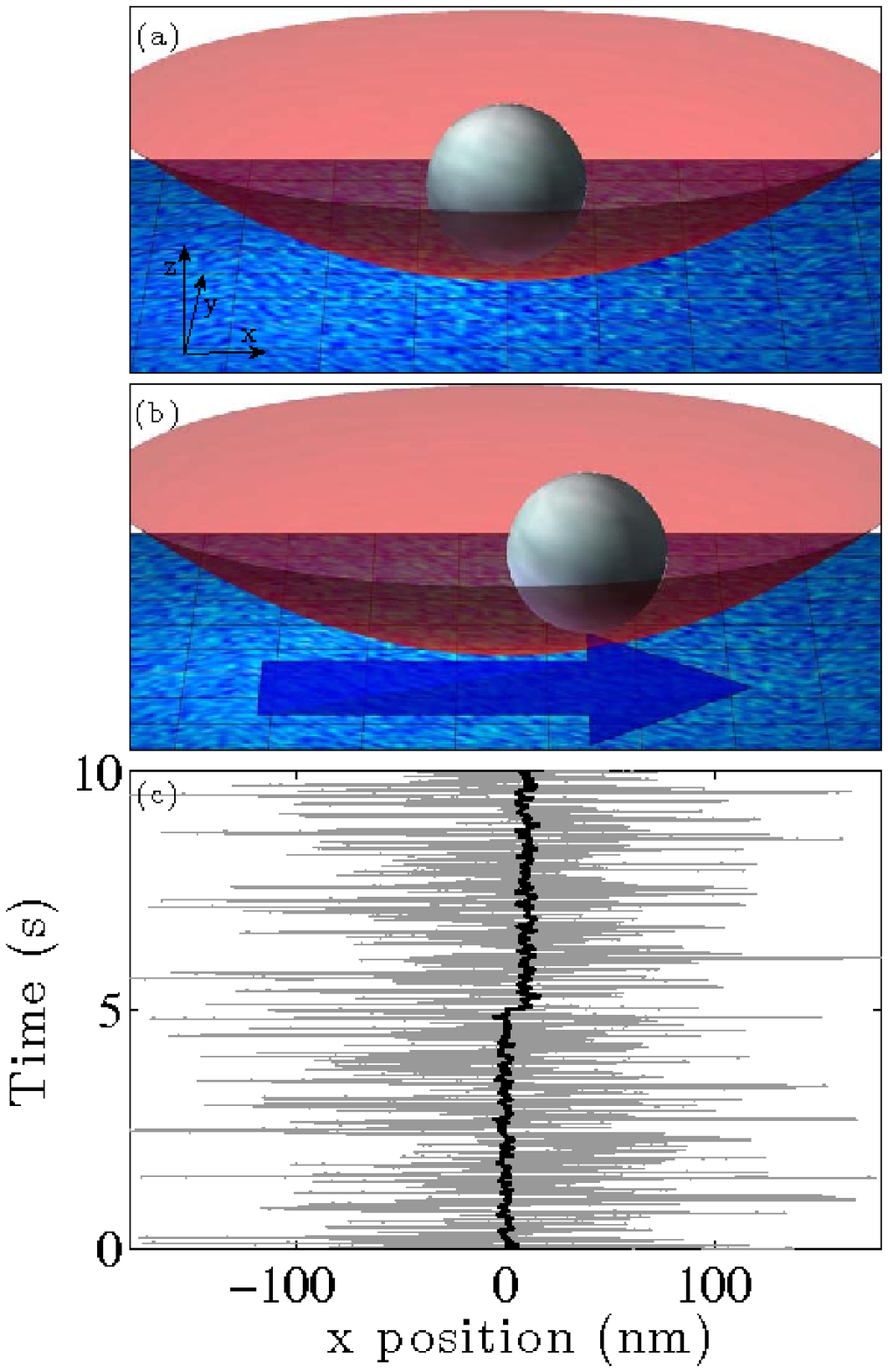}
\caption{Schematic of the experiment: (a) a mesoscopic particle performs
Brownian motion in the harmonic potential well generated by an optical
tweezers; (b) at time $t=5\,\mathrm{s}$ the action of a fluid flow 
exerts a force on the particle and displaces it from its equilibrium position.
(c) A single trajectory (grey) and the average over 400
trajectories  (black); the measured average displacement resulted equal to $10\pm6$ nm.
\label{fig1}}
\end{figure}

\begin{figure}
\includegraphics[width=14cm]{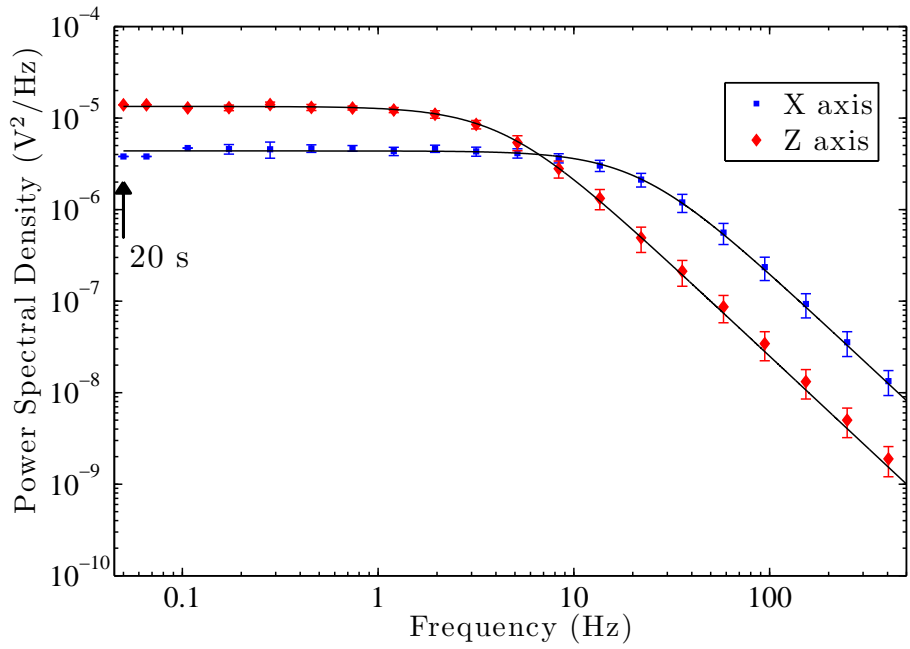}
\caption{Stability of the experimental setup. Experimental
power spectral densities, which are calculated from the measured
particle trajectories \cite{Berg-Sorensen2004} for the $x$ (blue squares) and $z$ (red
diamonds) coordinates, are well fitted by Lorentzians functions (solid
lines) in the range of interest for our experiments, i.e. from $50\,
\mathrm{mHz}$ to $500\, \mathrm{Hz}$. 
\label{fig2}}
\end{figure}

The duration of each trajectory measurement is $10\, s$ (figure
\ref{fig1}(c)). During the first 5 seconds the stage is at rest and the
particle is in an equilibrium state in the optical potential well,
$U(\rho) = k_{\rho} \rho^2/2$ (figure \ref{fig1}(a)), where $\rho$ is the radial position
of the Brownian particle with respect to the center of the trap, i.e. $\rho = \sqrt{x^2+y^2}$, and
$k_{\rho}$ is the radial stiffness of the optical trap, assumed to be harmonic.  At $t=5\,\mathrm{s}$ the stage starts moving
at a speed $v=1\, \mathrm{\mu m/s}$ along the $x$ coordinate (figure \ref{fig1}(b)),
producing an effective fluid flow and an effective
force $F = \gamma v$ acting on the particle, where $\gamma$ is the
particle friction coefficient.  Hence, the manipulation protocol is
\begin{equation} 
F(t) = 
\left\{ 
\begin{array}{cc} 
0 & t<5\, \mathrm{s} \\ 
\gamma v & t>5\,\mathrm{s}
\end{array}
\right.
\label{eq:protocol}
\end{equation}
After a pause of $1\,\mathrm{s}$ the above described sequence
starts again, but the stage is moved in the opposite direction.  This
allows us to remain in the same region on the sample cell avoiding
systematic drifts. By repeating this procedure 400 times, we acquire 800
trajectories for each particle.

During the application of the force, the average trajectory of the
particle (black solid line in figure \ref{fig1}(c)) climbs the optical
potential; however, due to the overwhelming presence of the Brownian
motion, each realization (e.g., gray solid line in figure \ref{fig1}(c))
presents a random behavior.

Assuming a harmonic trapping potential, as the stage moves with
constant velocity $v$ along the $x$ direction, the work done on the
Brownian particle reads \cite{Imparato2007}
\begin{equation} W(t)=\int_{t_0}^t ds\, v\, k_x x(s),
\label{eq:theo_work}
\end{equation} corresponding to the classical expression
``displacement times force," where $ds\, v$ is the infinitesimal
displacement of the center of the potential and $k_x x(s)$ is the
force acting on the particle with $k_x$ the trap stiffness along the $x$ direction.

In the case of an ideal Gaussian trap, the work probability
distribution is given by \cite{spe_seif_05, Imparato2007}
\begin{equation} \Phi(W,t) = \frac{1}{\sqrt{2 \pi \sigma(t)^2}}
\exp\pq{ -\frac{\left(W - \mu(t)\right)^2}{2\sigma(t)^2} },
\label{pw}
\end{equation} which is a Gaussian whose mean and variance depend on
time, i.e. $\mu(t) = v^2\tau_x^2 k_x \left(e^{-t/\tau_x} -1 + t/\tau_x
\right)$ and $\sigma(t)^2 = 2 \pi k_B T \mu(t)$ with $\tau_x =
\gamma/k_x$ the relaxation time in the optical trap, $T$ the
temperature and $k_B$ the Boltzmann constant.  As will be shown, a
deviation from a harmonic trap, leads to deviations from such work
distribution.  Thus, in the presence of nonconservative optical
forces, we expect the experimental probability distribution function
of the work to deviate from eq.~\ref{pw}.

\section{Work measurement on driven Brownian particles}

From the measured trajectories we calculated the work performed by the
fluid flow on a Brownian particle according to eq.~\ref{eq:theo_work}. 
The results are plotted as histograms in figure
\ref{fig3} for Si (first row), Ps (second row) and Me (third row)
particles and for $t = 10\, \mathrm{ms}$ (first column), $t=500\, \mathrm{ms}$ (second
column) and $t = 3000\, \mathrm{ms}$ (third column). The theoretical
expectation for a conservative force field according to eq.~\ref{eq:theo_work} are plotted as solid
lines. For $t=10\, \mathrm{ms}$, the work performed on all particles is in good
agreement with the theoretical expectation. However, for larger times
only the data relative to the Si particle keep on being in
agreement, while deviations arise for the data on Ps and
Me particles.

\section{Influence of rotational force fields}

In order to understand such disagreement, we consider the deviation of
the optical force field from the harmonic one assumed in
eq.~\ref{eq:theo_work}. 
Indeed, in the case of optical trapping powers as small as the ones we employ, 
the optical force field can be better approximated allowing for a rotational component in 
the vertical plane \cite{Pesce2009EPL}. 

A generic force field can be characterized by measuring the auto-correlation and cross-correlation 
functions between the coordinates of the particle \cite{PhysRevLett.97.210603,PhysRevE.76.061118,
PhysRevE.77.037301}. 
In particular, here in order to characterize the rotational optical forces, we use the 
cross correlation difference between $\rho$ and $z$ \cite{Pesce2009EPL}
\begin{eqnarray} 
\mathcal{C}_{\rho z} (\tau) = 
2\frac{k_B T}{\gamma} \frac{\epsilon}{1+k_z/k_{\rho}} 
\exp \left[  - \frac{|\tau|}{\bar{\tau}} \right]
\frac{\sinh \left[  \sqrt{| \bar{\tau}^{-2} - \epsilon^2 \tau_{\rho}^{-2} |} \tau \right] }{ \sqrt{| \bar{\tau}^{-2} - \epsilon^2 \tau_{\rho}^{-2} |} }
\label{eq:CCF}
\end{eqnarray} 
where $\bar{\tau} = \frac{1}{2}[\tau_{\rho} + \tau_z]$ is the average relaxation time in the trap, 
$\tau_z = \gamma/k_z$ the relaxation time along the $z$-direction, $k_z$ the stiffness along the 
$z$-direction and $\epsilon$ the relative contribution of the nonconservative force fields.
$\mathcal{C}_{\rho z} (\tau) \cong 0$ is clear evidence that the contribution of the 
nonconservative force field component is effectively negligible.

In figure \ref{fig4} the experimental $\mathcal{C}_{\rho z}
(\tau)$ is presented for the case of a Si (blue), Ps
(red) and Me (green) optically trapped Brownian particle without
fluid flow. While the rotational component is negligible for the case
of a Si particle, it becomes significant for the cases of Ps and,
even more, for the case of a Me particle.  Summarizing, the rotational
component appears to increase with the refractive index of the
particle.  The same results are obtained in the presence of a fluid
flow.

\begin{figure}
\includegraphics[width=14cm]{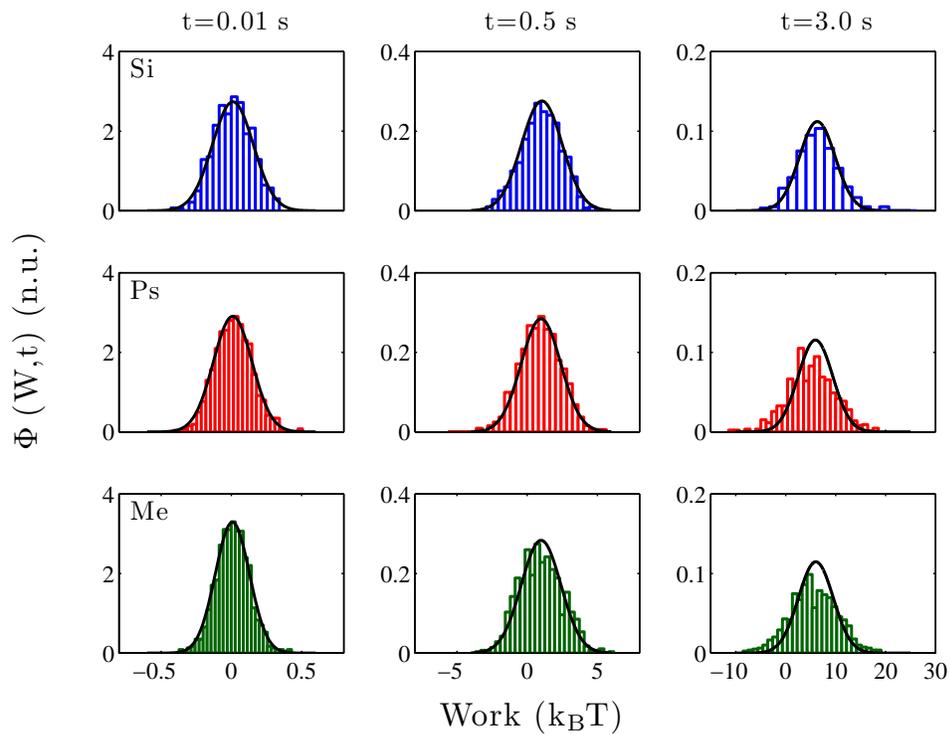}
\caption{Work distributions. Work performed on a silica (first row),
polystyrene (second row) and melamine (third row) Brownian particle driven by a
fluid flow for three different times from the beginning of the
manipulation protocol. Experimental data (histograms) and the
theoretical expectation (lines). Deviations are present for long times
in the case of polystyrene and melamine.
\label{fig3}}
\end{figure}

\begin{figure}
\includegraphics[width=14cm]{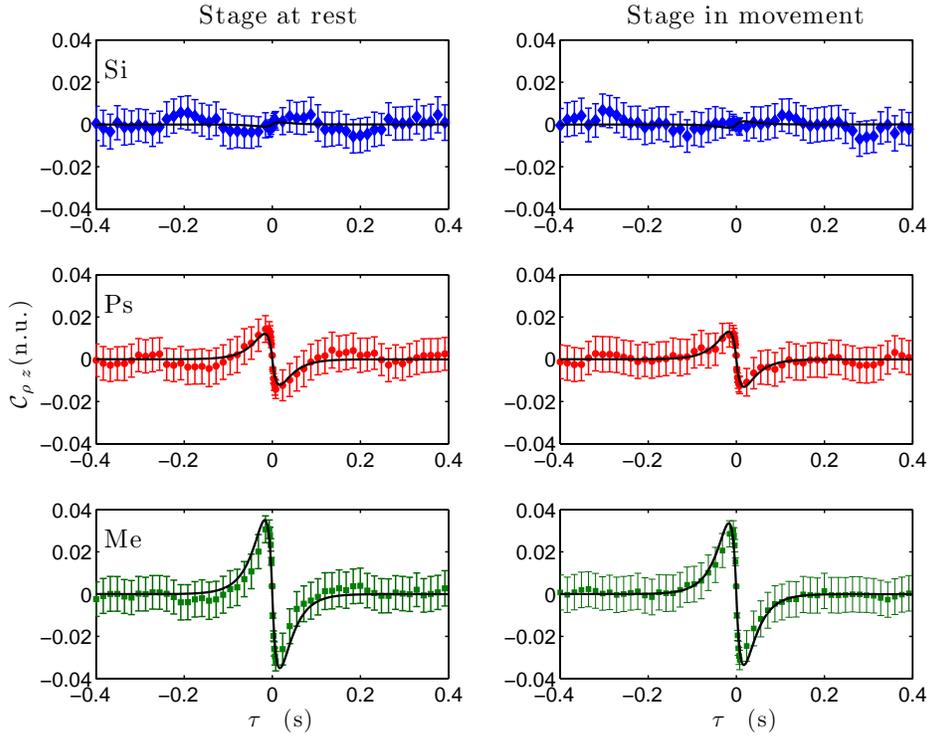}
\caption{Rotational radiation forces. Cross-correlation functions for
the $\rho$ and $z$ coordinates of a silica (first row), polystyrene (second row) and
melamine (third row) optically trapped Brownian particle with the fluid at
rest (left) or in the presence of a fluid flow (right).
\label{fig4}}
\end{figure}

The presence of a larger rotational component of the force field
appears to be in direct relation to the deterioration of the agreement
between theory and experiments in the determination of the work done
by the fluid flow on the Brownian particle. This can be understood in
a qualitative way.  In the presence of a harmonic trapping potential,
the particle movement along $x$, $y$ and $z$ are independent. For
example, assuming the particle to be at $x(t=0)=0$ the protocol acts
on the particle trajectory independently from the value of $z(t=0)$. 
However, if there is a rotational component of
the force field, a particle will undergo the effect of both the
protocol and the rotational force field, which, e.g., may result in a longer displacement for
$z(t=0)>0$ and in a shorter one for $z(t=0)<0$. Clearly, this
effect becomes evident only for times long enough for the effect of
the rotational component to be significant. This explains the widening
of the experimental work distribution in the case of Ps and Me particles for
long times (figure \ref{fig3}).

In principle, the partial differential equation ruling the time
evolution of the probability distribution function of the work could be
obtained also in the presence of nonconservative forces following the
procedure outlined in \cite{Imparato2007}. However, while such an
equation can be explicitly solved in the case of a harmonic potential,
leading to eq.~(\ref{pw}), this is not as easy for the case of
nonconservative forces, because it involves a nonlinear
problem. The discrepancy between the experimental data and the
theoretical prediction eq.~(\ref{pw}) observed in figure \ref{fig3} can
thus be attributed to the fact that eq.~(\ref{eq:theo_work}), and thus eq.~(\ref{pw}), only takes into account the contribution of
the conservative forces to the work.

\section{Conclusions and outlook}

We have shown how the presence of rotational radiation forces due to a
weak optical trap affects the measurement of the work distributions on
a driven particle. As we had already observed in \cite{Pesce2009EPL},
the rotational radiation forces become significative only for a very
low trapping power, i.e. below $1\,\mathrm{mW}$.
This is the regime in which various experiments to test thermodynamics relations on colloidal systems have been performed \cite{Wang2002, Imparato2007}.
We have now shown how
such spurious effects increase with an increasing refractive index
difference between the particle and the medium and that they become
undetectable in the case of a Si particle trapped in water even for
an extremely low trapping power. These observations will prove useful to minimize
the influence of rotational force fields in future experiments.

\newpage

\newpage

\newpage

\newpage

\end{document}